\documentclass{article}
\usepackage{amsmath, amsthm}

\usepackage{graphicx}
\newtheoremstyle{theorem}
  {10pt}		  
  {10pt}  
  {\sl}  
  {\parindent}     
  {\bf}  
  {. }    
  { }    
  {}     
\theoremstyle{theorem}

\newtheoremstyle{defi}
  {10pt}		  
  {10pt}  
  {\rm}  
  {\parindent}     
  {\bf}  
  {. }    
  { }    
  {}     
\theoremstyle{defi}

\begin{document}

\title{Theoretical construction of Morris-Thorne wormholes compatible with quantum field theory}

\author{Peter K.F. Kuhfittig\\
Department of Mathematics\\
Milwaukee School of Engineering\\
Milwaukee, WI  53202-3109 USA\\
kuhfitti@msoe.edu\\[2pt]}
\date{}

\maketitle
\begin{abstract}\noindent
This paper completes and extends some earlier studies by the 
author to show that Morris-Thorne wormholes are compatible with 
quantum field theory.  The strategy is to strike a balance between 
reducing the size of the unavoidable exotic region and the degree 
of fine-tuning of the metric coefficients required to achieve this 
reduction, while simultaneously satisfying the constraints from 
quantum field theory.  The fine-tuning also serves to satisfy 
various traversability criteria such as tidal constraints and 
proper distances through the wormhole.  The degree of fine-tuning 
turns out to be a generic feature of the type of wormhole 
discussed.

\phantom{a}
\noindent
PAC numbers: 04.20.Jb, 04.20.Gz


\end{abstract}

\phantom{a}
\section{Introduction}
Wormholes are handles or tunnels in the spacetime topology linking 
two separate and distinct regions of spacetime.  These regions may 
be part of our Universe or of different universes altogether.  The 
pioneer work of Morris and Thorne \cite{MT88} has shown that 
macroscopic wormholes may be actual physical objects.  Furthermore, 
such wormholes require the use of exotic matter to prevent 
self-collapse.  Such matter is confined to a small region around 
the throat, a region in which the weak energy condition is violated.  
Since exotic matter is rather problematical, it is desirable to 
keep this region as small as possible.  However, the use of 
arbitrarily small amounts of exotic matter leads to severe problems, 
as discussed by Fewster and Roman \cite{FR05a, FR05b}.  The 
discovery by Ford and Roman \cite{FR95,FR96} that quantum field 
theory places severe constraints on the wormhole geometries has 
shown that most of the ``classical" wormholes could not exist on 
a macroscopic scale.  The wormholes described by Kuhfittig 
\cite{pK06, pK08} are earlier attempts to strike a balance between two 
conflicting requirements, reducing the amount of exotic matter and 
fine-tuning the values of certain parameters.  The purpose of this 
paper is to refine and solidify the earlier ideas, particularly the 
use of the quantum inequalities of Ford and Roman, here slightly extended, 
all with the aim of demonstating 
that wormholes, which are based on Einstein's theory, are 
compatible with quantum field theory.  The models discussed will 
therefore (1) satisfy all the constraints imposed by quantum field 
theory, (2) strike a reasonable balance between a small proper 
thickness of the exotic region and the degree of fine-tuning of the 
metric coefficients, Eq.~(\ref{E:line}) below, (3) minimize the 
assumptions on these metric coefficients, and (4) satisfy certain 
traversabilty criteria.    

Problems with arbitrarily small amounts of exotic matter are also 
discussed in Ref. \cite{oZ07}, but the author states explicitly 
that the issues discussed here and in Ref. \cite{pK06} are beyond 
the scope of his paper.  

\section{The problem}
Consider the general line element
\begin{equation}\label{E:line}
   ds^2 =-e^{2\beta(r)}dt^2+e^{2\alpha(r)}dr^2+r^2(d\theta^2+
      \text{sin}^2\theta\, d\phi^2),
\end{equation}
where $\beta(r)\rightarrow 0$ and $\alpha(r)\rightarrow 0$ as 
$r\rightarrow \infty$.  (We are using units in which $G=c=1$.)  The 
function $\beta$ is called the \emph{redshift function}; this function 
must be everywhere finite to avoid an event horizon at the throat.  
The function $\alpha$ is related to the \emph{shape function} 
$b=b(r)$: 
\begin{equation}\label{E:line2}
   e^{2\alpha(r)}=\frac{1}{1-b(r)/r}.
\end{equation}
(The shape function determines the spatial shape of the wormhole when 
viewed, for example, in an embedding diagram.)  It now follows that 
\begin{equation}\label{E:shape}
    b(r)=r(1-e^{-2\alpha(r)}).
\end{equation}
The minimum radius $r=r_0$ is the \emph{throat} of the wormhole, 
where $b(r_0)=r_0$.  As a result, $\alpha$ has a vertical 
asymptote at the throat $r=r_0$:
\begin{equation}\label{E:asymptote} 
   \lim_{r \to r_0+}\alpha(r)=+\infty.
\end{equation}
So $\alpha(r)$ is assumed to be monotone decreasing near the throat.  
The qualitative features (again near the throat) of 
$\alpha(r)$, $\beta(r)$, and $-\beta(r)$, the reflection of 
$\beta(r)$ in the horizontal axis, are shown in Fig.~1.  It is 
assumed that $\beta$ and $\alpha$ are twice differentiable with 
$\beta'(r)\ge 0$ and $\alpha'(r)<0$.
\begin{figure}[htbp]
\begin{center}
\includegraphics[clip=true, draft=false, bb=0 0 299 212, 
  angle=0, width=4.24in, height=3in, 
   viewport=50 50 296 200]{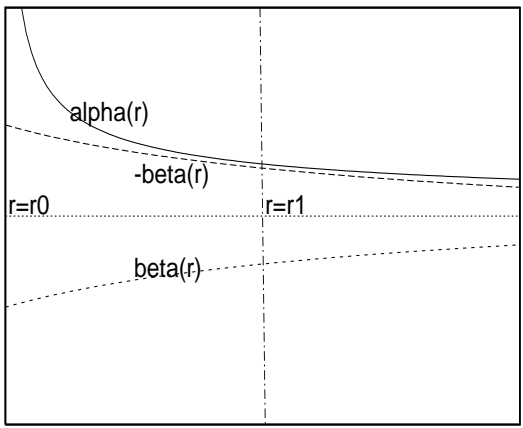}
\end{center}
\caption{\label{fig:figure1}Qualitative features of $\alpha(r)$ 
   and $\beta(r)$ near the throat.}
\end{figure} 

The next step is to list the components of the Einstein tensor in the 
orthonormal frame.  From Ref.~\cite{pK08},
\begin{equation}\label{E:E1}
   G_{\hat{t}\hat{t}}=\frac{2}{r}e^{-2\alpha(r)}\alpha'(r)
   +\frac{1}{r^2}(1-e^{-2\alpha(r)}),
\end{equation}
\begin{equation}\label{E:E2}
    G_{\hat{r}\hat{r}}=\frac{2}{r}e^{-2\alpha(r)}\beta'(r)
   -\frac{1}{r^2}(1-e^{-2\alpha(r)}),
\end{equation}
and
\begin{equation}\label{E:E3}
   G_{\hat{\theta}\hat{\theta}}=G_{\hat{\phi}\hat{\phi}}=
    e^{-2\alpha(r)}\left[\beta''(r)
      +\alpha'(r)\beta'(r)
     +[\beta'(r)]^2+\frac{1}{r}\beta'(r)
      -\frac{1}{r}\alpha'(r)\right].   
\end{equation}
Now recall that since the Einstein field equations 
$G_{\hat{\alpha}\hat{\beta}}=8\pi T_{\hat{\alpha}\hat{\beta}}$ 
in the orthonormal frame imply that the stress-energy tensor 
is proportional to the Einstein tensor, the only nonzero 
components are $T_{\hat{t}\hat{t}}=\rho,$ 
$T_{\hat{r}\hat{r}}=-\tau,$ and 
$T_{\hat{\theta}\hat{\theta}}=T_{\hat{\phi}\hat{\phi}}=p,$
where $\rho$ is the energy density, $\tau$ the radial tension, 
and $p$ the lateral pressure.  The weak energy condition (WEC) 
may now be stated as follows: the stress-energy tensor 
$T_{\hat{\alpha}\hat{\beta}}$ must obey 
\[
     T_{\hat{\alpha}\hat{\beta}}\mu^{\hat{\alpha}}\mu^{\hat{\beta}}\ge0  
\]
for all time-like vectors and, by continuity, all null vectors. 
Using the radial outgoing null vector $\mu^{\hat{\alpha}}=(1,1,0,0)$, 
the condition becomes $T_{\hat{t}\hat{t}}+T_{\hat{r}\hat{r}}=
\rho-\tau\ge 0.$  So if the WEC is violated, then $\rho-\tau<0$.
The field equations $G_{\hat{\alpha}\hat{\beta}}=
8\pi T_{\hat{\alpha}\hat{\beta}}$ now imply that 
\begin{equation}\label{E:WEC}
  \rho-\tau=\frac{1}{8\pi}\left(\frac{2}{r}e^{-2\alpha(r)}
    \left[\alpha'(r)+\beta'(r)\right]\right).
\end{equation}
Sufficiently close to the asymptote, $\alpha'(r)+\beta'(r)$ is 
clearly negative.  (Recall that $\alpha'<0$ and $\beta'\ge 0$.)
According to Ford and Roman \cite{FR95, FR96}, the exotic matter 
must be confined to a thin band around the throat.  To satisfy 
these constraints, we would like the WEC to be satisfied outside 
of some small interval $[r_0,r_1]$.  In other words, 
\begin{equation}\label{E:FR1}
    |\alpha'(r_1)|=\beta'(r_1),
\end{equation}
\begin{equation}\label{E:FR2}
    \alpha'(r)+\beta'(r)<0\quad \text{for}\quad r_0<r<r_1,
\end{equation}
and
\begin{equation}\label{E:FR3}
    \alpha'(r)+\beta'(r)\ge 0 \quad \text{for} \quad r\ge r_1.
\end{equation}
(See Fig.~1.)  Condition (\ref{E:FR3}) implies that 
$|\alpha'(r)|\le \beta'(r)$ for $r\ge r_1$.  So if 
$\beta(r)\equiv$ constant, then $\alpha'(r)\equiv0$ for $r\ge 
r_1$.  In the neighborhood of $r=r_1$, we also require that 
$\alpha''(r)>0$, $\beta''(r)<0$, and $\alpha''(r)>|\beta''(r)|$. 
We now have the minimum requirements for constructing the type 
of wormhole that we are interested in.

Using the components of the stress-energy tensor allows us to 
restate Eqs. (\ref{E:E1}) and (\ref{E:E2}) in terms of $b=b(r)$:
\begin{equation}\label{E:stress1}
    T_{\hat{t}\hat{t}}=\rho=\frac{b'(r)}{8\pi r^2}
\end{equation} 
and
\begin{equation}\label{E:stress2}
   T_{\hat{r}\hat{r}}=-\tau
   =-\frac{1}{8\pi}\left[\frac{b(r)}{r^3}-\frac{2\beta'(r)}{r}
    \left(1-\frac{b(r)}{r}\right)\right].
\end{equation}
Because of Eq. (\ref{E:stress1}), we require that $b'(r)>0$.  
Eq. (\ref{E:line2}) implies that $\alpha(r)=-\frac{1}{2}
\text{ln}(1-b(r)/r)$.  From
\begin{equation}\label{E:alphaprime}
  \alpha'(r)=\frac{1}{2}\frac{1}{1-b(r)/r}\frac{b'(r)-b(r)/r}{r},
\end{equation}
we conclude that $b'(r_0)\le 1$ to keep $\alpha'(r)$ negative 
near the throat.  In fact, $\lim_{r \to r_0+}\alpha'(r)=-\infty$.  
(The condition $b'(r_0)\le 1$ is called the \emph{flare-out} 
condition in Ref. \cite{MT88}.)

\subsection{The extended quantum inequality (first version)}\noindent
The sought-after compatibility with quantum field theory is based 
on the so-called quantum inequality in Ref. \cite{FR96}, applied 
to different situations.  (A modified version, based on Ref. 
\cite{FR05a}, is given in Sec. 6.)  This inequality deals 
with an inertial Minkowski spacetime without boundaries.  If 
$u^{\mu}$ is the observer's four-velocity (i.e., the tangent 
vector to a timelike geodesic), then 
$\langle T_{\mu\nu}u^{\mu}u^{\nu}\rangle$ is the expectation 
value of the local energy density in the observer's frame of 
reference.  It is shown that 
\begin{equation}\label{E:FR}
   \frac{\tau_0}{\pi}\int^{\infty}_{-\infty}
   \frac{\langle T_{\mu\nu}u^{\mu}u^{\nu}\rangle d\tau}
    {\tau^2+\tau_0^2}\ge -\frac{3}{32\pi^2\tau_0^4},
\end{equation}
where $\tau$ is the observer's proper time and $\tau_0$ the 
duration of the sampling time.  (See Ref.~\cite{FR96} for 
details.)  Put another way, the energy density is sampled in a 
time interval of duration $\tau_0$ which is centered around an 
arbitrary point on the observer's worldline so chosen that 
$\tau=0$ at this point.  It is shown in Ref. \cite{FR96} that the 
inequality can be applied in a curved spacetime as long as 
$\tau_0$ is small compared to the local proper radii of curvature, 
as illustrated in Ref. \cite {FR96} by several examples.  To 
obtain an estimate of the local curvature, we need to list the 
nonzero components of the Riemann curvature tensor in the 
orthonormal frame.  From Ref.~\cite{pK08}
\begin{equation}\label{E:Riemann1}
  R_{\hat{r}\hat{t}\hat{r}\hat{t}}=e^{-2\alpha(r)}
   \left(\beta''(r)-\alpha'(r)\beta'(r)
      +\left[\beta'(r)\right]^2\right),
\end{equation}
\begin{equation}\label{E:Riemann2}
   R_{\hat{\theta}\hat{t}\hat{\theta}\hat{t}}
    =R_{\hat{\phi}\hat{t}\hat{\phi}\hat{t}}
   =\frac{1}{r}
     e^{-2\alpha(r)}\beta'(r),
\end{equation}
\begin{equation}\label{E:Riemann3}
   R_{\hat{\theta}\hat{r}\hat{\theta}\hat{r}}
   = R_{\hat{\phi}\hat{r}\hat{\phi}\hat{r}}
   =\frac{1}{r}
       e^{-2\alpha(r)}\alpha'(r),      
\end{equation}
and
\begin{equation}\label{E:Riemann4}
   R_{\hat{\theta}\hat{\phi}\hat{\theta}\hat{\phi}}
   =\frac{1}{r^2}\left(1-e^{-2\alpha(r)}\right).     
\end{equation}
Still following Ref. \cite{FR96}, we need to introduce the 
following length scales over which various quantities change:
\begin{equation}\label{E:rsubm}
   r_m \equiv\text{min}\left[r, 
   \left|\frac{b(r)}{b'(r)}\right|,
      \frac{1}{|\beta'(r)|},
   \left|\frac{\beta'(r)}{\beta''(r)}\right|\right].
\end{equation}
The reason is that the above components of the Riemann 
curvature tensor can be reformulated as follows:
\begin{multline}\label{E:newRiemann1}
  R_{\hat{r}\hat{t}\hat{r}\hat{t}}=
   \left(1-\frac{b(r)}{r}\right)\frac{1}
    {\frac{\beta'(r)}{\beta''(r)}\frac{1}{\beta'(r)}}
    -\frac{b(r)}{2r}\left(\frac{1}
   {\frac{1}{\beta'(r)}\frac{b(r)}{b'(r)}} 
    -\frac{1}{r\frac{1}{\beta'(r)}}\right)\\
  +\left(1-\frac{b(r)}{r}\right)
     \frac{1}{\left(\frac{1}{\beta'(r)}\right)^2},
\end{multline}
\begin{equation}\label{E:newRiemann2}
   R_{\hat{\theta}\hat{t}\hat{\theta}\hat{t}}
    =R_{\hat{\phi}\hat{t}\hat{\phi}\hat{t}}
    =\left(1-\frac{b(r)}{r}\right)
  \frac{1}{r\frac{1}{\beta'(r)}},
\end{equation}
\begin{equation}\label{E:newRiemann3}
   R_{\hat{\theta}\hat{r}\hat{\theta}\hat{r}}
   = R_{\hat{\phi}\hat{r}\hat{\phi}\hat{r}}
   =\frac{b(r)}{2r}\left(
     \frac{1}{r\frac{b(r)}{b'(r)}}
   -\frac{1}{r^2}\right),      
\end{equation}
and
\begin{equation}\label{E:newRiemann4}
   R_{\hat{\theta}\hat{\phi}\hat{\theta}\hat{\phi}}
   =\frac{1}{r^2}\frac{b(r)}{r}.     
\end{equation}
When it comes to curvature, we are going to be primarily 
interested in magnitudes.  So we let $R_{\text{max}}$ denote 
the magnitude of the maximum curvature.  We know that the 
largest value of $(1-b(r)/r)$ and of $b(r)/r$ is unity; 
it follows from Eqs. (\ref{E:rsubm})-(\ref{E:newRiemann4}) 
that $R_{\text{max}}\le 1/r^2_m$ (disregarding the 
coefficient $\frac{1}{2}$).  So the smallest radius of 
curvature $r_c$ is 
\begin{equation}
   r_c\approx \frac{1}{\sqrt{R_{\text{max}}}}\ge r_m.
\end{equation}
The point is that working on this scale, the spacetime is 
Minkowskian (at least approximately), so that inequality 
(\ref{E:FR}) can be applied with an appropriate $\tau_0$.

As noted earlier, we assume that $b'(r)$ and hence $\rho$ 
are positive.  Being nonnegative, it is suggested in Ref. 
\cite{FR96} that a bound can be obtained by Lorentz 
transforming to the frame of a radially moving \emph{geodesic} 
observer who is moving with velocity $v$ relative to the 
static frame.  In this ``boosted frame"
\begin{equation}
   r'_c\approx \frac{1}{\sqrt{R'_{\text{max}}}}
   \ge \frac{r_m}{\gamma},
\end{equation} 
where $\gamma=(1-v^2)^{-1/2}$, so that the spacetime should be 
approximately flat.  The suggested sampling time is
\begin{equation}
   \tau_0=\frac{fr_m}{\gamma}\ll r'_c,
\end{equation}  
where $f$ is a scale factor such that $f\ll 1$.  The energy 
density in the boosted frame is 
\begin{equation}
   T_{\hat{0}'\hat{0}'}=\rho'=\gamma^2(\rho+v^2p_r),
\end{equation}
where $v$ is the velocity of the boosted observer.  It is 
stated in Ref. \cite {FR96} that in this frame the energy 
density does not change very much over the short sampling 
time and is therefore approximately constant:
\begin{multline}
   \frac{\tau_0}{\pi}\int^{\infty}_{-\infty}
   \frac{\langle T_{\mu\nu}u^{\mu}u^{\nu}\rangle d\tau}
    {\tau^2+\tau_0^2}\approx 
    \langle T_{\mu\nu}u^{\mu}u^{\nu}\rangle 
    \frac{\tau_0}{\pi}\int^{\infty}_{-\infty}
     \frac{d\tau}{\tau^2+\tau_0^2}\\
   =\langle T_{\mu\nu}u^{\mu}u^{\nu}\rangle=\rho' 
     \ge -\frac{3}{32\pi^2\tau_0^4}.
\end{multline}    
From Eqs. (\ref{E:stress1}) and (\ref{E:stress2}),
\[
   \rho'=\frac{\gamma^2}{8\pi r^2}\left[b'(r)
   -v^2\frac{b(r)}{r}+v^2r(2\beta'(r))
    \left(1-\frac{b(r)}{r}\right)\right].
\]
In order for $\rho'$ to be negative, $v$ has to be 
sufficiently large:
\begin{equation}\label{E:velocity}
  v^2>\frac{b'(r)}{\frac{b(r)}{r}-2r\beta'(r)
     \left(1-\frac{b(r)}{r}\right)};
\end{equation}
(observe that $v^2$ is dimensionless.)
In particular, at the throat, $v^2>b'(r_0)$.  Given $b(r)$, 
inequality (\ref{E:velocity}) places a restriction on 
$\beta'(r)$.  We will return to this point in Sec. 
\ref{S:solution}.

Next, from
\[
    \frac{3}{32\pi^2\tau_0^4}\ge -\rho'
\]
we have
\begin{equation*}
  \frac{32\pi^2\tau_0^4}{3}\le\\\frac{8\pi r^2}{\gamma^2}
   \left[v^2\frac{b(r)}{r}-b'(r)-v^2r(2\beta'(r))
    \left(1-\frac{b(r)}{r}\right)\right]^{-1}.
\end{equation*}
Using $\tau_0=fr_m/\gamma$ and dividing both sides by 
$r^4$, we have (disregarding a small coefficient)
\begin{equation*}
    \frac{f^4r_m^4}{r^4\gamma^4} \le\\\frac{1}{r^2\gamma^2}
    \left[v^2\frac{b(r)}{r}-b'(r)-2v^2r\beta'(r)
        \left(1-\frac{b(r)}{r}\right)\right]^{-1}
\end{equation*}
and, after inserting $l_p$ to produce a dimensionless 
quantity,
\begin{equation}\label{E:genQI}
   \frac{r_m}{r}\le 
    \left(\frac{1}{v^2\frac{b(r)}{r}-b'(r)-2v^2r\beta'(r)
    \left(1-\frac{b(r)}{r}\right)}\right)^{1/4}\\
       \frac{\sqrt{\gamma}}{f}
        \left(\frac{l_p}{r}\right)^{1/2}.
\end{equation}
This is the first version of the extended quantum inequality.  
At the throat, where $b(r_0)=r_0$, inequality (\ref{E:genQI}) 
reduces to Eq. (95) in Ref. \cite{FR96}:
\begin{equation}\label{E:QI}
  \frac{r_m}{r_0}\le\left(\frac{1}{v^2-b'(r_0)}\right)^{1/4}
       \frac{\sqrt{\gamma}}{f}\left(\frac{l_p}{r_0}\right)^{1/2}.
\end{equation}
Observe that inequality (\ref{E:QI}) is trivially satisfied if 
$b'(r_0)=1$ but not if $b'(r_0)<1$.  In view of inequality 
(\ref{E:genQI}) and the tidal constraints in the next 
subsection, we would like $b'(r)$ to be close to unity in 
the exotic region.  (The need for $b'(r_0)$ to be close to 1 
is also pointed out in Ref. \cite{FR96}.) 

\subsection{The tidal constraints} 

Much of what follows is based on the discussion in 
Ref. \cite{MT88}.  In particular, we have for the radial tidal 
constraint 
\begin{multline}\label{E:radial}
  \left|R_{\hat{1}'\hat{0}'\hat{1}'\hat{0}'}\right|=
  \left|R_{\hat{r}\hat{t}\hat{r}\hat{t}}\right|\\
    =e^{-2\alpha(r)}
      \left|\beta''(r)-\alpha'(r)\beta'(r)
      +\left[\beta'(r)\right]^2\right|
     \le \frac{g_{\oplus}}{c^2\times 2\,\text{m}}
   \approx (10^8\,\text{m})^{-2},
\end{multline}
that is, assuming a traveler with a height of 2 m.  This 
constraint is trivially satisfied if $\beta(r)\equiv$ 
constant, referred to as the zero-tidal-force solution in 
Ref. \cite{MT88}.
The lateral tidal constraints are (reinserting $c$)
\begin{multline}\label{E:lateral}
  \left|R_{\hat{2}'\hat{0}'\hat{2}'\hat{0}'}\right|
  =\left|R_{\hat{3}'\hat{0}'\hat{3}'\hat{0}'}\right|
  =\gamma^2\left|R_{\hat{\theta}\hat{t}\hat{\theta}\hat{t}}\right|
  +\gamma^2\left(\frac{v}{c}\right)^2\left|
     R_{\hat{\theta}\hat{r}\hat{\theta}\hat{r}}\right|\\
   =\gamma^2\left(\frac{1}{r}e^{-2\alpha(r)}\beta'(r)\right)
   +\gamma^2\left(\frac{v}{c}\right)^2
       \left(\frac{1}{r}e^{-2\alpha(r)}\alpha'(r)\right)
     \le (10^8\,\text{m})^{-2};
\end{multline} 
here $\gamma^2=1/\left[1-(v/c)^2\right]$.

Returning to Eq.~(\ref{E:shape}), 
we have for the shape function,
\begin{equation}\label{E:bprime}
    b'(r_0)=\frac{d}{dr}\left[r(1-e^{-2\alpha(r)})\right]_{r=r_0}\\
    =2r_0e^{-2\alpha(r_0)}\alpha'(r_0)+1-e^{-2\alpha(r_0)}.
\end{equation}
In order for $b'(r_0)\approx 1,$ we require that 
\[
   \lim_{r \to r_0+}e^{-2\alpha(r)}\alpha'(r)=0.
\]
As a consequence, the radial tidal constraint (\ref{E:radial}) 
is satisfied at the throat, while the lateral tidal constraints 
(\ref{E:lateral}) merely constrain the velocity of the traveler
in the vicinity of the throat.

One of the consequences of the condition $b'(r_0)\approx 1$ is
 that the wormhole will flare out very slowly, so that the 
coordinate distance from $r=r_0$ to $r=r_1$ will be much less 
than the proper distance.  

\subsection{The exotic region}
We saw in the last section that $\alpha$ has to go to infinity fast 
enough so that $\lim_{r \to r_0+}e^{-2\alpha(r)}\alpha'(r)=0.$  At 
the same time, $\alpha$ has to go to infinity slowly enough so that 
the proper distance  
\begin{equation*}
   \ell(r)=\int\nolimits_{r_0}^{r}e^{\alpha(r')}dr'
\end{equation*}
is finite.  Then by the mean-value theorem, there exists a value 
$r=r_2$ such that
\[
   \ell(r)=e^{\alpha(r_2)}(r-r_0),\quad r_0<r_2<r.
\]
In particular, $\ell(r_0)=0$ and 
\begin{equation}\label{E:meanvalue}
  \ell(r_1)=e^{\alpha(r_2)}(r_1-r_0).
\end{equation}
With this information we can examine the radial tidal constraint 
at $r=r_1.$  From Eq.~(\ref{E:Riemann1})
\begin{multline*}\label{E:radial1}
  |R_{\hat{r}\hat{t}\hat{r}\hat{t}}|=e^{-2\alpha(r_1)}
   \left|\beta''(r_1)-\alpha'(r_1)\beta'(r_1)
      +\left[\beta'(r_1)\right]^2\right|\\
     =e^{-2\alpha(r_1)}\left|\beta''(r_1)
       -\alpha'(r_1)[-\alpha'(r_1)]
         +\left[\alpha'(r_1)\right]^2\right|
\end{multline*}
by Eq.~(\ref{E:FR1}).  So by inequality (\ref{E:radial}), 
\begin{equation*}
 |R_{\hat{r}\hat{t}\hat{r}\hat{t}}|
    =e^{-2\alpha(r_1)}\left|\beta''(r_1)+
     \alpha'(r_1)\alpha'(r_1)
         +\left[\alpha'(r_1)\right]^2\right|\\
            \le (10^8\text{m})^{-2}.
\end{equation*}
Since $e^{-2\alpha(r)}$ is strictly increasing, it follows that
\[
     e^{-2\alpha(r_2)}\left|\beta''(r_1)
       +2\left[\alpha'(r_1)\right]^2\right|
            <10^{-16} \text{m}^{-2}.
\]
From Eq.~(\ref{E:meanvalue}), we now get the following:
\[
   \frac{(r_1-r_0)^2}{[\ell(r_1)]^2}
    \left|\beta''(r_1)+2\left[\alpha'(r_1)\right]^2\right|
            <10^{-16}\text{m}^{-2}
\]
and
\begin{equation}\label{E:abs1}
  \left|\beta''(r_1)+2\left[\alpha'(r_1)\right]^2\right|
  <\frac{[\ell(r_1)]^2}{10^{16}(r_1-r_0)^2}.
\end{equation}
As a consequence,
\begin{equation}\label{E:abs2}
  \beta''(r_1)+2\left[\alpha'(r_1)\right]^2
   <\frac{[\ell(r_1)]^2}{10^{16}(r_1-r_0)^2}
\end{equation}
or
\begin{equation}\label{E:abs3}
  \beta''(r_1)+2\left[\alpha'(r_1)\right]^2
   >-\frac{[\ell(r_1)]^2}{10^{16}(r_1-r_0)^2}.
\end{equation}
So if either condition (\ref{E:abs2}) or condition (\ref{E:abs3})
is satisfied, then so is condition (\ref{E:abs1}).

\section{A class of models; fine-tuning}
To estimate the size of the exotic region, we need some 
idea of the magnitude of $\alpha(r)$, which depends on the 
specific model chosen.  The only information available is 
that $\alpha(r)$ increases slowly enough as $r\rightarrow 
r_0+$ to keep $\int\nolimits_{r_0}^{r}e^{\alpha(r')}dr'$ 
finite.  (We have not made any assumptions regarding \
$\beta(r)$, except for some of the basic requirements.)

First we need to recall that Morris-Thorne wormholes are not 
just concerned with traversability in general but more 
specifically with humanoid travelers.  According to 
Ref.~\cite{MT88}, the space station should be far enough 
away from the throat so that 
\begin{equation}\label{E:flat}
   1-\frac{b(r)}{r}=e^{-2\alpha(r)}\approx 1,
\end{equation}
making the space nearly flat.  Another condition involves the 
redshift function: at the station we must also have
\begin{equation}\label{E:station1}
   |\beta'(r)|\le g_{\oplus}/\left(c^2\sqrt{1-b(r)/r}\right).
\end{equation}
It will be seen below that for our wormhole, the first 
condition, Eq.~(\ref{E:flat}), is easily satisfied.  By 
condition (\ref{E:FR3}), as well as Fig. 1, 
$|\alpha'(r)|\le \beta'(r)$ for $r\ge r_1$.  So if 
$1-b(r)/r\approx 1,$ then we have 
\begin{equation}\label{E:station2}
    |\alpha'(r)|<10^{-16}\,\text{m}^{-1}
\end{equation}  
at the station.  This inequality should give us at least a rough 
estimate of the distance to the station: for large $r$,
$|\alpha(r)|\sim|\beta(r)|$, since both $\alpha$ and $\beta$ 
go to zero.  It must be kept in mind, however, that the 
inequality $|\alpha'(r)|\le \beta'(r)$ implies that this 
procedure does underestimate the distance.  The main reason 
for using $\alpha$ in the first place is to avoid making 
additional assumptions involving $\beta$.  Instead, $\beta$ 
can be left to its more obvious role, helping to meet the 
tidal constraints and the quantum inequality, Eq. 
(\ref{E:genQI}).  We will return to this point 
after discussing $\alpha.$  

Consider next a class of models based on the following set of 
functions:
\begin{equation}\label{E:hyper}
  \alpha(r)=a\,\text{ln}\left(\frac{1}{(r-r_0)^b}
    +\sqrt{\frac{1}{(r-r_0)^{nb}}+1}\right).
\end{equation}
(Other models are discussed in Ref. (\cite{pK08}).)  For 
convenience let us concentrate for now on the special case 
$n=2$ and return to Eq.~(\ref{E:hyper}) later.  For $n=2$, 
the equation becomes
\begin{equation}\label{E:sinh}
    \alpha(r)=a\,\text{sinh}^{-1}\frac{1}{(r-r_0)^b},
        \quad b>\frac{1}{2a}.
\end{equation}
The need for the assumption $b>1/(2a)$ comes from the shape 
function 
\[
  b(r)=r\left(1-e^{-2a\,\text{sinh}^{-1}
   [1/(r-r_0)^b]}\right):
\]
\begin{multline*}
  b'(r)=1-e^{-2a\,\text{sinh}^{-1}[1/(r-r_0)^b]}\\
   +r\left(-e^{-2a\,\text{sinh}^{-1}[1/(r-r_0)^b]}\right)
    \frac{2ab}{(r-r_0)\sqrt{(r-r_0)^{2b}+1}};
\end{multline*}
$b'(r)\rightarrow 1$ as $r\rightarrow r_0$, as long as 
$b>1/(2a).$  To see this, it is sufficient to examine 
\[
  e^{-2a\,\text{sinh}^{-1}[1/(r-r_0)^b]}\frac{1}{r-r_0}
\]
as $r\rightarrow r_0$:
\begin{multline*}
  \frac{1}{\left[\frac{1}{(r-r_0)^b}+\sqrt{\frac{1}{(r-r_0)^{2b}}+1}
    \right]^{2a}}\frac{1}{r-r_0}\\
   =\frac{1}{\frac{1}{(r-r_0)^{2ab}}\left[1+(r-r_0)^b
    \sqrt{\frac{1}{(r-r_0)^{2b}}+1}\right]^{2a}}\frac{1}{r-r_0}\\
    =\frac{1}{\frac{1}{(r-r_0)^{2ab-1}}
    \left[1+\sqrt{1+(r-r_0)^{2b}}\right]^{2a}}.  
\end{multline*}
So if $2ab-1>0$, then the second factor in the denominator 
becomes negligible for $r\approx r_0$.  The result is
\[
   e^{-2a\,\text{sinh}^{-1}[1/(r-r_0)^b]}\frac{1}{r-r_0}
   \sim(r-r_0)^{2ab-1}\rightarrow 0.
\]
For computational purposes, however, we will simply let $b=1/(2a)$.  
Consider next,
\begin{equation}\label{E:D1alpha}
  \alpha'(r)=-\frac{ab}{(r-r_0)\sqrt{(r-r_0)^{2b}+1}},\quad r>r_0,
\end{equation}
and
\begin{equation}\label{E:D2alpha}
   \alpha''(r)=\frac{ab\left[(1+b)(r-r_0)^{2b}+1\right]}
      {(r-r_0)^2\left[(r-r_0)^{2b}+1\right]^{3/2}}.
\end{equation}

We know that the wormhole flares out very slowly at the throat, 
which suggests assigning a small coordinate distance to the exotic 
region, at least initially.  A good choice is $r-r_0=0.000001$ m, as 
in Ref. \cite{pK08}.  Then from Eqs. (\ref{E:D1alpha}) and 
(\ref{E:D2alpha}), we get
\begin{equation}
    \alpha'(r_1)\approx -\frac{ab}{r_1-r_0}\quad \text{and}
  \quad \alpha''(r_1)\approx\frac{ab}{(r_1-r_0)^2}.
\end{equation}
For future reference, let us replace $ab$ by $A$:
\begin{equation}\label{E:alphageneral}
  \alpha'(r_1)\approx -\frac{A}{r_1-r_0}\quad \text{and}
  \quad \alpha''(r_1)\approx \frac{A}{(r_1-r_0)^2}.
\end{equation}
Since we also want $\alpha''(r_1)>|\beta''(r_1)|$ [or
$\alpha''(r_1)>-\beta''(r_1)$], we have in view of 
inequality (\ref{E:abs2}),  
\begin{equation}\label{E:main}
  \frac{A}{(r_1-r_0)^2}>-\beta''(r_1)>\frac{2A^2}{(r_1-r_0)^2}-
     \frac{[\ell(r_1)]^2}{10^{16}(r_1-r_0)^2}.  
\end{equation}
Conversely, the inequality 
\begin{equation*}
   \frac{2A^2}{(r_1-r_0)^2}-\frac{[\ell(r_1)]^2}
          {10^{16}(r_1-r_0)^2}\\
    =2[\alpha'(r_1)]^2-\frac{[\ell(r_1)]^2}{10^{16}(r_1-r_0)^2}
    <-\beta''(r_1)  
\end{equation*}
implies condition (\ref{E:abs2}).  Since $-\beta''(r_1)<
\alpha''(r_1)$, we conclude that inequality (\ref{E:main}) 
is valid if, and only if, condition (\ref{E:abs2}) is met.

Inequality (\ref{E:main}) now implies that 
\begin{equation}
   2A^2-A-\frac{[\ell(r_1)]^2}{10^{16}}<0.
\end{equation}
The critical values are
\[
    A=\frac{1\pm\sqrt{1+\frac{8[\ell(r_1)]^2}{10^{16}}}}{4}.
\]
Hence
\[
 \frac{1-\sqrt{1+\frac{8[\ell(r_1)]^2}{10^{16}}}}{4}<A<
 \frac{1+\sqrt{1+\frac{8[\ell(r_1)]^2}{10^{16}}}}{4}.
\]
Returning to the condition $b'(r_0)\le 1$ for a moment, note
that $ab$ and hence $A$ must exceed 1/2.  It follows that 
\begin{equation}\label{E:Afinal}
   \frac{1}{2}<A<\frac{1+\sqrt{1+\frac{8[\ell(r_1)]^2}{10^{16}}}}{4}
\end{equation}
and, replacing $A$,
\begin{equation}\label{E:abfinal}
  \frac{1}{2}<ab<\frac{1+\sqrt{1+\frac{8[\ell(r_1)]^2}{10^{16}}}}{4}.   
\end{equation}
The left inequality confirms that $b>1/(2a)$.
This solution shows that considerable fine-tuning is required.  We 
will return to this point in Sec.~\ref{S:finetune}.

Finally, observe that with the extra condition $|\beta''(r_1)|
<\alpha''(r_1)$, the  qualitative features in Fig.~1 are retained,
so that no additional assumptions are needed.

Letting $b=1/(2a)$ once again for computational purposes, we now have 
\[
   \ell(r_1)=\int_{r_0}^{r_0+0.000001}
   e^{a\,\text{sinh}^{-1}[1/(r-r_0)^{1/(2a)}]}dr.
\]
These values change very little with $a$.  For example, if $a$ 
ranges from 0.1 to 0.5, then $\ell(r_1)$ ranges from 0.0021 m 
to 0.0028 m; $\ell(r_1)$ is much larger than $r_1-r_0$, a 
consequence of the slow flaring out.   From inequality 
(\ref{E:station2}) we can estimate the distance $r_s$ to 
the space station: if 
$|\alpha'(r_s)|=10^{-16}\,\text{m}^{-1}$, then 
$r_s=70\,000$ km. Of course, we can always reduce the 
coordinate distance.  Thus for 
$r_1-r_0=0.000000001\,\text{m}$ and $a=0.5$, we get 
$\ell(r_1)=0.000089\,\text{m}<0.1\,\text{mm}$.

A good alternative is to use Eq.~(\ref{E:hyper}), subject to 
the condition
\[
    nab-b+\frac{1}{2}nb>1.
\]
(As before, this condition comes from the requirement that 
$b'(r_0)\le 1$; in fact, if $n=2,$ we are back to $2ab>1$.)  
For example, retaining $r_1-r_0=0.000001\,\text{m}$, if 
$a=0.2$ and $b=1$, then $nb=2.857$.  These values yield 
$\ell(r_1)\approx 0.0000725\,\text{m}<0.1\,\text{mm}.$  
The corresponding distance $r_s$, obtained from 
$\alpha'(r)$ [now referring to Eq.~(\ref{E:hyper})], 
is about $45\,000$ km.  Both $\ell(r_1)$ and $r_s$ are 
relatively small.
 
Using the equation $nab-b+\frac{1}{2}nb=1$ to eliminate $n$ 
in Eq.~(\ref{E:hyper}) shows that further reductions in 
$\ell(r_1)$ are only significant if $a$ and $b$ get 
unrealistically close to zero.  So practically speaking, a  
further reduction in the proper distance $\ell(r_1)$ 
requires a reduction in the coordinate distance $r_1-r_0$.

Returning to the radial tidal constraint, based on experience 
with specific functions (as in Ref.~\cite{pK06}), 
$\left|R_{\hat{r}\hat{t}\hat{r}\hat{t}}\right|$ is likely to 
reach its peak just to the right of $r=r_1$.  The simplest 
way to handle this problem is to 
tighten the constraint in Eq.~(\ref{E:radial}) at $r=r_1$ by 
reducing the right side.  This change increases the degree of 
fine-tuning in condition (\ref{E:Afinal}).  

A final consideration is the time dilation near the throat. 
Denoting the proper distance by $\ell$ and the proper time by 
$\tau$, as usual, we let $\gamma v=d\ell/d\tau$, so that 
$d\tau=d\ell/(\gamma v)$.  Assume that $\gamma\approx 1$.  
Since $d\ell=e^{\alpha(r)}dr$ and $d\tau=e^{\beta(r)}dt$, 
we have for any coordinate time interval $\Delta t$:
\begin{equation*}
  \Delta t=\int\nolimits_{t_a}^{t_b}dt=
     \int\nolimits_{\ell_a}^{\ell_b}e^{-\beta(r)}\frac{d\ell}{v}= 
     \int\nolimits_{r_a}^{r_b}\frac{1}{v}e^{-\beta(r)}e^{\alpha(r)}
        dr.     
\end{equation*}
From Eq. (\ref{E:hyper}), we have on the interval $[r_0, r_1]$ 
\begin{equation*}
   \Delta t= \int_{r_0}^{r_1}\frac{1}{v}
    e^{-\beta(r)}\left(\frac{1}{(r-r_0)^b}
     +\sqrt{\frac{1}{(r-r_0)^{nb}}+1}\right)^adr.
\end{equation*}
Since $\beta(r)$ is finite, the small size of the interval 
$[r_0,r_1]$ implies that $\Delta t$ is going to be 
relatively small for a wide variety of choices for 
$a$ and $b$.

\section{The fine-tuning problem in general}\label{S:finetune}
The forms of inequalities (\ref{E:Afinal}) and 
(\ref{E:abfinal}) suggest that the degree of fine-tuning 
encountered is a general property of the type of wormhole 
being considered, namely wormholes for which $b'(r_0)\le 1$ and 
$\alpha(r)=A\,\text{ln}f(r-r_0),$ where (generalizing from 
earlier cases) $f(r-r_0)|_{r=r_0}$ is undefined $(+\infty)$ 
and $f(\frac{1}{r-r_0})|_{r=r_0}$ is a constant (possible 
zero).  If we also assume that $g(r-r_0)=f(\frac{1}{r-r_0})$
can be expanded in a Maclaurin series, then we have for 
$r\approx r_0$, 
\begin{multline*}
  f\left(\frac{1}{r-r_0}\right)=g(r-r_0)=a_0+a_1(r-r_0)\\
      +a_2(r-r_0)^2+a_3(r-r_0)^3+\cdot\cdot\cdot
     \approx a_0+a_1(r-r_0).
\end{multline*}  
It follows that 
\[
    f(r-r_0)=a_0+\frac{a_1}{r-r_0}
\]
near the throat.  So
\[
  \alpha(r)=A\,\text{ln}\left(a_0+\frac{a_1}{r-r_0}\right),
\]
\begin{equation}\label{E:der1}
   \alpha'(r_1)=\frac{-Aa_1}{a_0+\frac{a_1}{r_1-r_0}}
     \frac{1}{(r_1-r_0)^2}\sim -\frac{A}{r_1-r_0},
\end{equation}
and
\begin{equation}\label{E:der2}
   \alpha''(r_1)\sim\frac{A}{(r_1-r_0)^2}.
\end{equation}

To show that $b'(r_0)\le 1,$ we need to show that 
$e^{-2\alpha(r)}\alpha'(r)\rightarrow 0$ as $r\rightarrow 
r_0$:
\begin{multline*}
   e^{-2A\,\text{ln}[a_0+a_1/(r-r_0)]}\frac{-Aa_1}
      {a_0+\frac{a_1}{r-r_0}}\frac{1}{(r-r_0)^2}\\
    =\frac{1}{\left(a_0+\frac{a_1}{r-r_0}\right)^{2A}}
     \frac{-Aa_1}{a_0+\frac{a_1}{r-r_0}}
        \frac{1}{(r-r_0)^2}
   =\frac{-Aa_1}{\left(a_0+\frac{a_1}{r-r_0}\right)^{2A+1}
       (r-r_0)^2}\\
   =\frac{-Aa_1}{\left[\left(a_0+\frac{a_1}{r-r_0}\right)
       (r-r_0)\right]^{2A+1}\frac{(r-r_0)^2}{(r-r_0)^{2A+1}}}
   =\frac{-Aa_1}{[a_0(r-r_0)+a_1]^{2A+1}
         \frac{1}{(r-r_0)^{2A-1}}}.               
\end{multline*}
The first factor in the denominator becomes negligle for 
$r\approx r_0$ as long as $2A-1>0$ and $A>\frac{1}{2}.$  
We obtain 
\[
    e^{-2\alpha(r)}\alpha'(r)\sim(r-r_0)^{2A-1}
     \rightarrow 0.
\]
Comparing Eqs. (\ref{E:der1}) and (\ref{E:der2}) to 
Eq.~(\ref{E:alphageneral}), we conclude that
\begin{equation}\label{E:finetuning}
    \frac{1}{2}<A<\frac{1+\sqrt{1+\frac{8[\ell(r_1)]^2}
         {10^{16}}}}{4}.
\end{equation}
So the amount of fine-tuning required appears to be a 
general property of wormholes of the present type. (Exactly 
which parameter needs fine-tuning depends on the precise 
form of $f(r-r_0)$.) While the degree of fine-tuning 
considered so far is quite severe, it is considerably 
milder than most of the cases discussed in 
Ref.~\cite{FR05a}.

\section{The solution}\label{S:solution}
The discussion of Morris-Thorne wormholes in Ref. \cite{MT88} 
is concerned not just with traversability but, more 
specifically, with traversability by humanoid travelers.  So 
the length of the trip, possible time dilations, and the tidal 
constraints are important considerations.

The first part of this paper deals with the size of the 
unavoidable exotic region around the throat.  It was found 
that the size can be reduced almost indefinitely by carefully 
fine-tuning $\alpha=\alpha(r)$ or, equivalently, the shape 
function $b=b(r)$.  The degree of fine-tuning required of 
some parameter turns out to be a general property of the type 
of wormhole considered.  To achieve this fine-tuning, it is 
necessary to assume that $b'(r)$ is close to unity near the 
throat.  This assumption proved to be sufficient to satisfy 
the tidal constraints.

Concerning the quantum inequalities, if $b'(r_0)=1$, then 
inequality (\ref{E:QI}) is trivially satisfied at or near 
the throat.  Away from the throat that may not be the case.  
Fortunately, we have made no assumptions on $\beta=\beta(r)$ 
beyond the basic requirements, no event horizon and 
$\beta'(r)\ge |\alpha'(r)|$ for $r\ge r_1$.  For convenience, 
we restate inequalities (\ref{E:velocity}) and 
(\ref{E:genQI}),
\begin{equation}\label{E:velocity2}
  v^2>\frac{b'(r)}{\frac{b(r)}{r}-2r\beta'(r)
     \left(1-\frac{b(r)}{r}\right)},
\end{equation}
 
\begin{equation}\label{E:last}
   \frac{r_m}{r}\le 
    \left(\frac{1}{v^2\frac{b(r)}{r}-b'(r)-2v^2r\beta'(r)
    \left(1-\frac{b(r)}{r}\right)}\right)^{1/4}\\
      \frac{\sqrt{\gamma}}{f}
        \left(\frac{l_p}{r}\right)^{1/2},
\end{equation}
where $v$ is the velocity of the radially moving geodesic 
observer.  Since $\beta'(r)>0$, it now becomes evident 
that $\beta'(r)$ can be adjusted (or constructed ``by 
hand") to become part of the fine-tuning strategy:
according to Fig. 2, for a typical shape function 
\begin{figure}[htbp]
\begin{center}
\includegraphics[clip=true, draft=false, bb=0 0 299 212, 
   angle=0, width=4.5in, height=3.0in, 
   viewport=50 50 296 200]{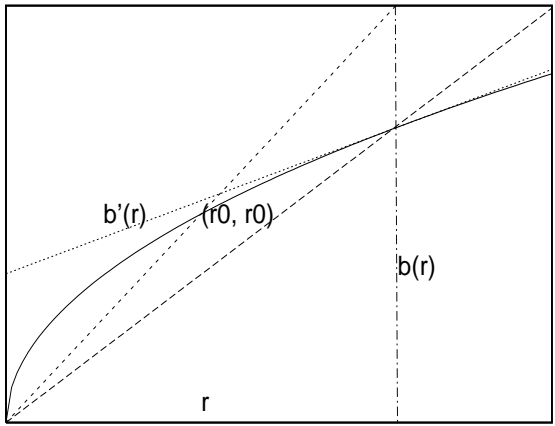}
\end{center}
\caption{\label{fig:figure2}Graph showing that 
   $\frac{b(r)}{r}-b'(r)>0$.}
\end{figure} 
$b=b(r)$, the slope of the tangent line at $(r, b(r))$ 
is less than the slope $b(r)/r$ of the chord extending 
from the origin to $(r, b(r))$.  This allows us to 
construct (or adjust) $\beta'(r)$ so that
\begin{equation}\label{E:construct}
  \frac{b(r)}{r}-b'(r)-2r\beta'(r)\left(1-\frac{b(r)}
      {r}\right)
\end{equation}
is 0 or very nearly 0.  (According to Eq. 
(\ref{E:alphaprime}), $\beta'(r)$ is large enough for 
$r>r_1$.)  Hence the right-hand side of inequality 
(\ref{E:velocity2}) is 1 or very nearly 1, thereby 
forcing $v$ to be 1 or very nearly 1.  As a consequence, 
the denominator on the right-hand side of inequality 
(\ref{E:last}) is 0 or very nearly 0; so the inequality is 
satisfied for any $r_m$.

\emph{Remark:}  As noted earlier, at $r=r_0$, inequality 
(\ref{E:last}) is trivially satisfied.  Similarly, at 
$r=r_1$, expression (\ref{E:construct}) is zero since 
$\beta'(r_1)=|\alpha'(r_1)|$.  To the right of $r_1$, 
$\beta'(r)$ is large enough to overtake $b(r)/r-b'(r)$ 
and can therefore be adjusted to produce 0 or very 
nearly 0.  Inside the small interval $[r_0,r_1]$, 
however, it may be necessary to fine-tune $b(r)$ to 
keep $b'(r)$ close to 1 inside the interval, or, which 
amounts to the same thing, $\alpha(r)$ must turn sharply 
upward after crossing $r=r_1$ from the right.  (Recall 
the qualitative features in Fig. 1.)  

Observe that, given any particular $b=b(r)$,
the choice $\beta \equiv$ constant is not 
likely to work, basically in agreement with the analysis 
in Ref. \cite{FR96}, since the original wormhole models 
in Ref. \cite{MT88} all assumed a constant $\beta$, at 
least near the throat. 

Since inequality (\ref{E:last}) is satisfied, the radius 
of the throat, $r=r_0$, is macroscopic since $r_m$ 
includes $r_0$.  The wormholes 
satisfy the various traversability criteria for humanoid 
travelers.  All the while the exotic region is made as 
small as possible while keeping the degree of fine-tuning 
within reasonable bounds.  The models discussed have led 
to the following promising results: approximately 0.1 mm 
for the proper thickness of the exotic region, corresponding 
to a distance much less than $100\,000$ km to the space 
station.  By decreasing the coordinate distance, it is 
theoretically possible to decrease the proper thickness 
of the exotic region indefinitely.  While the decrease 
may be thought of as an engineering challenge, the fact 
remains that the concomitant increase in the degree of 
fine-tuning would eventually exceed any practical limit.  

\section{Additional remarks: the extended quantum inequality 
         (second version)}\noindent
The extended quantum inequality discussed in Subsection 2.1 is not 
the most general form available: a version of the original quantum 
inequality based on the violation of the null energy condition is 
obtained in Ref. \cite{FR05a}.  This inequality, about to be extended, 
eliminates the need for a boosted frame since it features a static 
observer.

For present purposes it is sufficient to note that for the null vector 
$\textbf{k}=\textbf{e}_{\hat{t}}+\textbf{e}_{\hat{r}}$,
\[
   T_{\hat{\alpha}\hat{\beta}}k^{\hat{\alpha}}k^{\hat{\beta}}  
=\rho-\tau=-\frac{e^{2\beta(r)}}{8\pi r}\frac{d}{dr}
  \left[e^{-2\beta(r)}\left(1-\frac{b(r)}{r}\right)\right],
\]
which is readily obtained from Eq. (\ref{E:WEC}).  It follows 
from the subsequent discussion in Ref. \cite{FR05a} that 
\[
  \frac{e^{2\beta(r)}}{8\pi rl^2_p}\frac{d}{dr}
   \left[e^{-2\beta(r)}\left(1-\frac{b(r)}{r}\right)\right]
       \le \frac{C}{\tau^4_0}
\]
after inserting the Planck length $l_p$.  The constant $C$ is 
given in Eq. (7) of Ref. \cite{FR05a}.  It is assumed that 
$\tau_0=f\ell_{min}$, where $\ell_{min}$ is the proper minimum length 
scale.  Taking the derivative and solving for $\ell_{min}/r$, we obtain  
\begin{equation*}
   \frac{\ell_{min}}{r}\le 
    \left(\frac{1}{\frac{b(r)}{r}-b'(r)-2r\beta'(r)
    \left(1-\frac{b(r)}{r}\right)}\right)^{1/4}\\
      \frac{1}{f}\left(\frac{l_p}{r}\right)^{1/2}
         (8\pi C)^{1/4}.
\end{equation*}
According to Ref. \cite{FR05a}, $(8\pi C)^{1/4}\approx 3.2$.  Since 
we are primarily interested in estimating orders of magnitude, we now 
have inequality (\ref{E:last}) in Sec. 5 with $v$ and $\gamma$ 
omitted, while $\ell_{min}$ replaces $r_m$:  
\begin{equation}
   \frac{\ell_{min}}{r}\le 
    \left(\frac{1}{\frac{b(r)}{r}-b'(r)-2r\beta'(r)
    \left(1-\frac{b(r)}{r}\right)}\right)^{1/4}\\
      \frac{1}{f}\left(\frac{l_p}{r}\right)^{1/2}.
\end{equation}
As a result, our conclusions are unaltered.

\end{document}